\documentclass[aps,prb,twocolumn,superscriptaddress,showpacs,showkeys,floatfix]{revtex4-1}
\usepackage{graphicx}
\usepackage{amssymb}
\usepackage{setspace}
\usepackage{amsmath}
\usepackage{multirow}
\usepackage{csquotes}
\usepackage{float}
\usepackage {braket}
\usepackage{hyperref}
\usepackage{xcolor}
\usepackage{xfrac}
\usepackage{siunitx}
\usepackage{tabularx}
\usepackage{tabulary}
\usepackage{tabu}
\usepackage{mathtools}
\usepackage{xspace}
\usepackage{booktabs}
\usepackage{placeins}
\usepackage[title]{appendix}
\usepackage{titlesec}
\titlespacing{\section}
{0pt}{0.5ex plus 0.5ex minus .5ex}{0.5ex  plus 0.5ex minus .5ex}

\definecolor{dark-red}{rgb}{0.4,0.15,0.15}
\definecolor{dark-blue}{rgb}{0.15,0.15,0.4}
\definecolor{medium-blue}{rgb}{0,0,0.5}
\hypersetup{
    colorlinks, linkcolor={dark-red},
    citecolor={dark-blue}, urlcolor={medium-blue}
}

\newcommand{\fto}{FeTi$_2$O$_5$\xspace}
\newcommand{\cto}{CoTi$_2$O$_5$\xspace}
\newcommand{\musr}{$\mu^+$SR\xspace}
\newcommand{\RomanNumeralCaps}[1]
    {\MakeUppercase{\romannumeral #1}}

\begin{document}


\title{\fto: a spin Jahn-Teller transition enhanced by cation substitution}

\author{Franz~Lang}
\email{franz.lang@physics.ox.ac.uk}
\affiliation{Oxford University Department of Physics, Clarendon Laboratory, Parks Road, Oxford, OX1 3PU, United Kingdom}

\author{Lydia~Jowitt}
\affiliation{Oxford University Department of Physics, Clarendon Laboratory, Parks Road, Oxford, OX1 3PU, United Kingdom}

\author{Dharmalingam~Prabhakaran}
\affiliation{Oxford University Department of Physics, Clarendon Laboratory, Parks Road, Oxford, OX1 3PU, United Kingdom}

\author{Roger~D.~Johnson}
\affiliation{Oxford University Department of Physics, Clarendon Laboratory, Parks Road, Oxford, OX1 3PU, United Kingdom}

\author{Stephen~J.~Blundell}
\affiliation{Oxford University Department of Physics, Clarendon Laboratory, Parks Road, Oxford, OX1 3PU, United Kingdom}

\date{\today}

\begin{abstract}
We have used muon-spin rotation, heat capacity and x-ray diffraction measurements in combination with density functional theory and dipole field calculations to investigate the crystal and magnetic structure of \fto. We observe a long range ordered state below $T_{\rm N}$=41.8(5)~K with indications of significant correlations existing above this temperature. We determine candidate muon stopping sites in this compound, and find that our data are consistent with the spin Jahn-Teller driven antiferromagnetic ground state with $\boldsymbol{k}$=(1/2,1/2,0) reported for \cto ($T_{\rm N}$=26~K). By comparing our data with calculated dipolar fields we can restrict the possible moment size and directions of the Fe$^{2+}$ ions.
\end{abstract}


\maketitle

\section{Introduction\label{sec:intro}}
Compounds containing lattices of magnetic moments in which the exchange interactions are frustrated by the site symmetry of the moments are of great scientific interest due to the fact that comparably small terms in the Hamiltonian can play an important role in stabilising a particular ground state. This leads to a plethora of physical phenomena that can occur in such geometrically frustrated systems, such as spin ice states showing magnetic monopole excitations, spin liquids with fractional excitations, or topological spin textures. 
The Jahn-Teller effect is a spontaneous distortion of a systems that reduces its symmetry and energy, and lifts an orbital degeneracy. An analogous effect can occur that involves the magnetoelastic coupling of the spin instead of the orbital degrees of freedom. This spin Jahn-Teller effect is rather rare and until recently had been only identified in high symmetry scenarios. Examples include pyrochlore systems, in which the large degeneracy of the spins occupying a lattice of corner-sharing tetrahedra is lifted through a distortion of those tetrahedra~\cite{Yamashita2000,Tchernyshyov2002}. In a number of cubic spinels~\cite{Watanabe2012,Onoda2003} a tetragonal distortion has been reported to relax the frustration in an entirely analagous way. 

Recently, we reported an antiferromagnetic ground state in \cto~\cite{Kirschner2019}, which we concluded to be driven by a spin Jahn-Teller effect. We observe \cto to be orthorhombic and to undergo a transition to a long-range ordered state below 26~K with antiferromagnetically coupled chains of Co$^{2+}$ along the $a$ axis. As the symmetry of the structure dictates that the orbital levels of the Co ions are nondegenerate and the interchain couplings are exactly frustrated, the observed long-range ordered state necessitates the presence of a spin-driven lattice distortion.

One possible pathway to tuning the properties of such a delicate ground state is cation substitution. In this paper we report a structural and magnetic characterisation of \fto. High quality single crystals of \fto can now be prepared owing to recent advances in solid-state synthesis techniques~\cite{Balbashov2017}. This compound is isostructural to \cto~\cite{Muller-Buschbaum1983}, but importantly the Fe$^{2+}$ ($3d^6$) ions have larger ionic radii and increased moments compared to the Co$^{2+}$ ($3d^7$) ions, which will affect the competition between the structural and magnetic energies involved in reaching the magnetic ground state. We expect that the larger Fe moments will enhance the magnetostructural coupling, whereas the slightly larger wavefunctions of the Fe ions are likely to lead to larger overlap with the oxygen orbitals and thereby increase the superexchange (Fe--O--Fe) couplings. This leads to the interesting prospect that \fto may be more susceptible to structural distortions and that the effect of externally applied strain, either through chemical or physical means, would also be more significant, making it better suited to such studies.

In our measurements we observe a transition to a magnetically ordered  state below 41.8(5)~K, substantially higher than the transition temperature in \cto, and through a combination of calculational techniques demonstrate that the magnetic ground state is consistent with that found in \cto with $\boldsymbol{k}$=(1/2,1/2,0), with the Fe magnetic moment lying in one of a small number of possible directions, with moment sizes ranging from about 2.54~$\mu_{\rm B}$ to 3.44~$\mu_{\rm B}$ depending on the direction. Since the Fe$^{2+}$ ($3d^6$) ions have site symmetry $m2m$ and their super-superexchange is exactly frustrated this magnetic structure requires a structural distortion, which we expect to be concomittant with the onset of long-range spin order. This makes \fto another example demonstrating that the spin Jahn-Teller can occur in low symmetry compounds and such behaviour may even be universal in these orthorhombic systems.

\section{Experimental Details\label{sec:expdetails}}
A polycrystalline \fto sample was prepared using FeO (99.5\%)  and TiO$_2$ (99.99\%) powders via the solid state reaction technique. The powders were mixed and sintered at 1200\textdegree~C for 48~h in a highly pure argon atmosphere with intermediate grinding. We confirmed the resulting powder to be phase pure via x-ray diffraction, before sintering a cylindrical rod of diameter 10~mm and length 100~mm at 1250\textdegree~C in a highly pure argon atmosphere for 12~h. Finally, we grew a single crystal in a four-mirror optical floating-zone furnace (Crystal Systems, Inc.) in an argon flow atmosphere with a growth rate of 1--2~mm/h.

Single crystal x-ray diffraction experiments were performed using a laboratory based Oxford Diffraction Supernova diffractometer fitted with a Mo K$_\alpha$ x-ray source. The sample was kept at room temperature, and an approximately full sphere of data was collected giving 4805 statistically significant, measured diffraction intensities. The diffraction data were refined with \textsc{fullprof}\cite{Rodriguez-Carvajal1993}.

We performed zero-field (ZF) muon-spin rotation (\musr) measurements on a sample of five single crystals of \fto ($m$=0.265~g) in a $^4$He flow cryostat on the GPS instrument at the Swiss Muon Source, PSI (Switzerland). The crystals were coaligned with the $ab$-plane perpendicular to the beam of incoming muons, i.e. the crystals shared the $c$-axis, and the spin rotator was set to 45\textdegree\,relative to the muon momentum. All muon data were analysed with WiMDA~\cite{wimda}. The heat capacity was measured using a Quantum Design physical property measurement system (PPMS), while the magnetic susceptibility was measured with a Quantum Design magnetic property measurement system (MPMS).

\section{Results\label{sec:results}}
\subsection{Heat Capacity and Magnetic Susceptibility\label{sec:HC}}
The molar heat capacity of \fto, plotted in Figure~\ref{plot:AFFT}(a), exhibits a clear feature around 43~K. We obtain an estimate of the magnetic contribution $C_{\rm mag}$ to the heat capacity by subtracting a phenomenological phonon background term of the form $C_{\rm ph}$=$\gamma C_{\rm Debye}(\theta_{\rm D})+(1-\gamma)C_{\rm Einstein}(\theta_{\rm E})$, where the fitted values are $\theta_{\rm D}$=578~K, $\theta_{\rm E}$=42~K and $\gamma$=0.77. The result is presented in Figure~\ref{plot:AFFT}(b), together with the corresponding entropy $S_{\rm mag}$. The entropy associated with the magnetic transition is about 60\% of the expected value of Rln2, indicating significant correlations above $T_{\rm N}$. The magnetic susceptibility, shown in Figure~\ref{plot:AFFT}(c), is consistent with a magnetic transition near 43~K and a fit to a Curie-Weiss law, presented in Figure~\ref{plot:AFFT}(d), yields a Curie-Weiss temperature of $\theta_{\rm CW}$=$-145$(1)~K and hence a value of $|\theta_{\rm CW}|$/$T_{\rm N}$$\sim$3.5.

\subsection{X-ray Diffraction\label{sec:xray}}
A global fit of the unit cell metric showed that the x-ray diffraction peak positions were consistent with an orthorhombic space group, and the refined lattice parameters are given in Table \ref{TAB::structure}. Having indexed all measured intensities within the orthorhombic metric, data reduction using \textsc{fullprof}\cite{Rodriguez-Carvajal1993} showed the diffraction pattern to be in excellent agreement with the respective $mmm$ Laue Class ($R_\mathrm{int}$: 3.58\%), and gave 306 independent reflections for structural refinement. The $Cmcm$ orthorhombic crystal structure previously reported for FeTi$_2$O$_5$ and CoTi$_2$O$_5$ \cite{Muller-Buschbaum1983} was Rietveld-refined against the diffracton intensities, again using \textsc{fullprof}. An excellent agreement between model and data was achieved ($R_{F^2}=3.32\%$, $wR_{F^2}=4.41\%$, $R_F=1.72\%$). A statistically significant improvement to the fit was found when cation site mixing was included, constrained such that the total occupation of every cation site equalled one, but Fe was free to substitute Ti and vice versa. Approximately 10\% Fe was found to be located on nominally Ti sites, though this fraction is difficult to establish precisely as x-ray diffraction measurements are only weakly sensitive to the site mixing of cations with similar atomic number.
\begin{table}
\caption{\label{TAB::structure}Refined room temperature crystal structure parameters of FeTi$_2$O$_5$ ($R_{F^2}$=$3.32\%$, $wR_{F^2}$=$4.41\%$, $R_F$=$1.72\%$).}
\begin{ruledtabular}
\begin{tabular}{c l c c c c c}
\multicolumn{6}{l}{\textbf{Cell parameters}} \\
\multicolumn{6}{l}{Space group: $Cmcm$ (\#194)} \\
\multicolumn{6}{l}{$a,b,c$ ($\mathrm{\AA}$)  3.74098(9)  9.7609(2)  10.0914(2)} \\
\multicolumn{6}{l}{Volume ($\mathrm{\AA}^3$) 368.49(2)}\\ 
\multicolumn{6}{l}{\textbf{Atomic fractional coordinates}} \\
Atom & Site & $a$ & $b$ & $c$ & Occ.\\
\hline
Fe1 & $4c$ & 0 & 0.19322(6) & 1/4 & 0.896(9) \\
Ti1 & $4c$ & 0 & 0.19322(6) & 1/4 & 0.104(9) \\
Ti2 & $8f$ & 0 & 0.13401(5) & 0.56666(5) & 0.948(4) \\
Fe2 & $8f$ & 0 & 0.13401(5) & 0.56666(5) & 0.052(4) \\
O1\hphantom{1} & $4c$ & 0 & 0.7917(3)\hphantom{1} & 1/4 & 1 \\
O2\hphantom{1} & $8f$ & 0 & 0.0452(2)\hphantom{1} & 0.1123(2) & 1 \\
O3\hphantom{1} & $8f$ & 0 & 0.3148(2)\hphantom{1} & 0.0586(2)  & 1 \\
\multicolumn{6}{l}{\textbf{Anisotropic atomic displacement parameters} ($\mathrm{\AA}^2$)} \\
\multicolumn{6}{l}{(U$_{12}$ = U$_{13}$ = 0 for all sites)}\\
Atom && U$_{11}$ & U$_{22}$ & U$_{33}$ & U$_{23}$ & \\
\hline
Fe1 && 0.0062(3) & 0.0040(3) & 0.0061(3) & 0 \\
Ti2 && 0.0042(3) & 0.0048(3) & 0.0066(3) & 0.0003(2) \\
O1  && 0.005(1)\hphantom{1}  & 0.008(1)\hphantom{1}  & 0.005(1)\hphantom{1}  & 0 \\
O2  && 0.013(1)\hphantom{1}  & 0.0042(9) & 0.0050(9) & 0.0000(8) \\
O3  && 0.003(1)\hphantom{1} &  0.0059(9) & 0.008(1)\hphantom{1} & -0.0005(7) \\
\multicolumn{6}{l}{\textbf{Data reduction} ($R_\mathrm{int}$: 3.58\%)} \\
\multicolumn{6}{l}{$\#$ measured reflections: 4805} \\
\multicolumn{6}{l}{$\#$ independent reflections $(I > 1.5\sigma$): 306} \\
\multicolumn{6}{l}{$\#$ fitted parameters: 29} \\
\end{tabular}
\end{ruledtabular}
\end{table}

\subsection{\musr\label{sec:musr}}
\begin{figure}[htb] \center
\includegraphics[width=\columnwidth, clip, trim= 0.0mm 0.0mm 0.0mm 0.0mm]{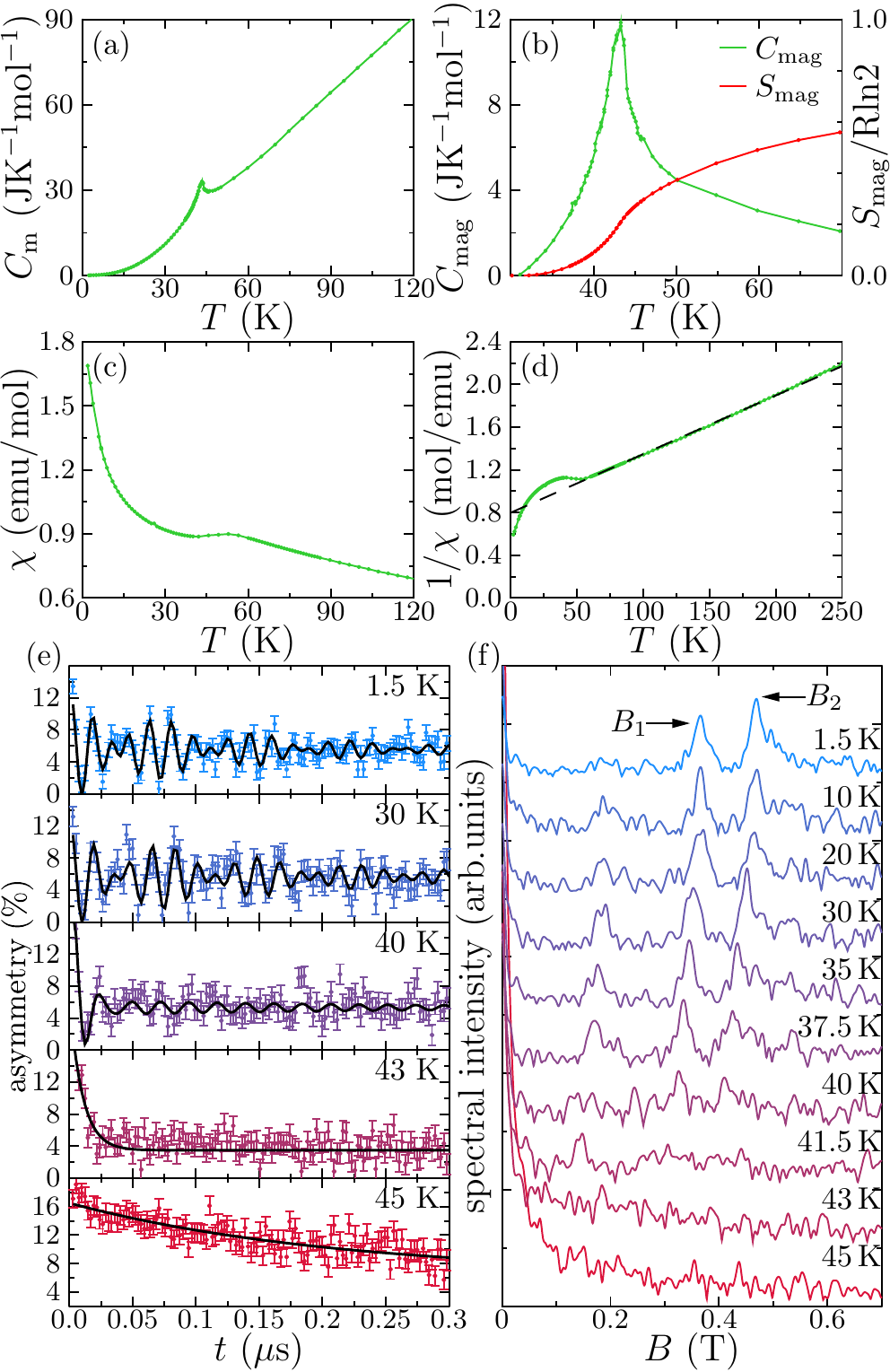}
\caption{ \label{plot:AFFT}
(a) Molar heat capacity $C_{\rm m}$. (b) Magnetic contribution to $C_{\rm m}$ and its associated entropy, obtained by subtracting a phonon background from $C_{\rm m}$ as described in the main body. (c) Magnetic susceptibility $\chi$ in a 50~Oe external field. (d) 1/$\chi$ with the dashed line representing a fit to a Curie-Weiss law, which yields a Curie-Weiss temperature of $\theta_{\rm CW}$=$-145$(1)~K. (e) ZF muon asymmetry $A(t)$ for a range of temperatures (vertically displaced for clarity). Solid lines represent fits to equation~\ref{eq:asym}. (f) Fourier transforms of $A(t)$ for a range of temperatures (vertically displaced for clarity). The two marked peaks correspond to the two fields fitted in equation~\ref{eq:asym}.
}
\end{figure}

\vspace{-12pt}
The measured ZF muon asymmetry $A(t)$ is shown in Figure~\ref{plot:AFFT} together with its Fourier transform. There is clear evidence for a transition to a state with long range magnetic order, and also for two symmetry inequivalent muon sites with different local magnetic fields. There is a possible third peak in the Fourier transform near 0.2~T, but this is not clearly resolvable for all $T$$<$$T_{\rm N}$, including for the highest statistics dataset at 1.5~K. Our time-domain fitting did not converge when including this third peak, so we discount it from our analysis. In general, the asymmetry can be well fitted for all $T$$<$$T_{\rm N}$ to the model
\begin{equation}
A(t)=A_{\rm r}\sum_{i=1,2}a_i\cos(\gamma_\mu B_i t)e^{-\lambda_i t}+A_{\rm b}
\label{eq:asym}
,\end{equation}
where $\gamma_\mu$=$2\pi$$\times$135.5~MHzT$^{-1}$ is the muon gyromagnetic ratio. The terms in the sum represent muons at two different stopping sites precessing in their local fields $B_i$, with the relaxation rates $\lambda_i$ allowing for temporal fluctuations of the fields. The background term $A_{\rm b}$ corresponds to muons that stop in the sample but have their spins aligned with the local field, as well as a small contribution from muons landing in the cryostat or sample holder. The values of $A_{\rm r}$ and $A_{\rm b}$ are approximately $T$-independent as they only depend on the total flux of muons and the setup of the experiment. Their temperature averaged values are $\langle$$A_r$$\rangle$=11.2(5)~\% and $\langle$$A_b$$\rangle$=5.3(3)~\%. The fraction $a_1$ ($a_2$) of muons experiencing field $B_1$ ($B_2$) is also roughly constant  with a value of 0.57(7) [0.43(7)]. Furthermore, we note that $\lambda_1$$\approx$$\lambda_2$, suggesting the two muon sites to be of similar character. Above $T_{\rm N}$ we can fit $A(t)$ by setting $B_1$=$B_2$=0 in equation~\ref{eq:asym}.

\begin{figure}[htb] 
\center
\includegraphics[width=\columnwidth, clip, trim= 0.0mm 0.0mm 0.0mm 0.0mm]{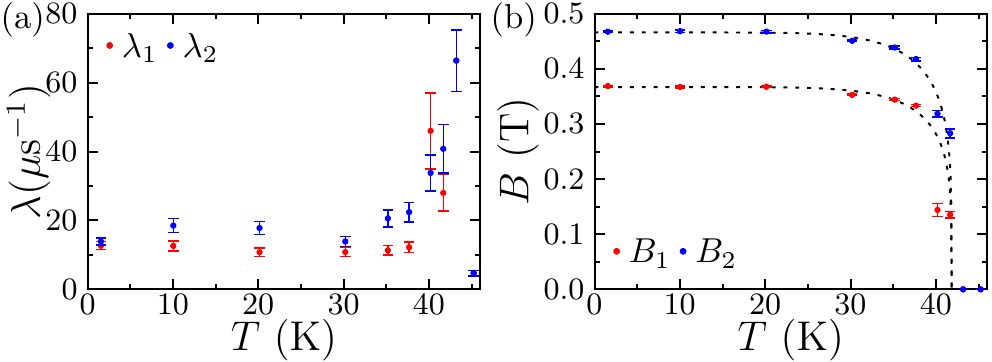}
\caption{ \label{plot:Afits}
Results of fitting equation~\ref{eq:asym} to the muon asymmetry $A(t)$. The dotted lines represent fits to a phenomenological order parameter model as described in the main body.
}
\end{figure}

The results of fitting equation~\ref{eq:asym} are plotted in Figure~\ref{plot:Afits}. The relaxation rates $\lambda_i$ peak at $T_{\rm N}$, consistent with a critical slowing down near the transition. The fields $B_i$ follow an order parameter like behaviour and can be modelled with the phenomenological form $B_i$=$B_i(0)(1-(T/T_{\rm c})^\alpha)^\beta$, from which we obtain $T_{\rm c}$=41.8(5)~K, as well as $B_1(0)$=0.367(4)~T and $B_2(0)$=0.466(5)~T. The fitted values of $\alpha$=7.5(3.5) and $\beta$=0.21(13) are unreliable to identify the dimensionality of the behaviour due to the low number of temperature points.

We expect the effect of the site mixing on the muon measurements to be limited as the volume fraction of muons near a mixed site will be below our experimental sensitivity and the likely effect to muon sites further away from a mixed site will be a broadening of the magnetic field spectrum. The values of $\lambda_i$ at low temperatures are approximately double those found for \cto,~\cite{Kirschner2019} due to a wider field distribution which we attribute to the site mixing.

\section{Muon Sites\label{sec:dft}}
To determine candidate muon sites in \fto we carried out density functional theory (DFT) calculations using the plane-wave code {\sc Quantum ESPRESSO}~\cite{Gianozzi2009,Giannozzi2017} within the generalised gradient approximation~\cite{Perdew1996}. Ultrasoft pseudopotentials were employed to model the ions~\cite{Rappe1990}, and a norm-conserving hydrogen pseudopotential was used for the muon. The cutoff for the wavefunction (charge density) was set to 60~Ry (600~Ry), and the Brillouin zone integration was carried out on a 5$\times$3$\times$3 Monkhorst-Pack $k$-grid~\cite{Monkhorst1976}. Within these parameters we obtained well-converged results that reproduced the atomic positions and lattice parameters within 3.5\% of the experimental ones~\cite{Muller-Buschbaum1983}. 

\begin{figure}[htb] 
\center
\includegraphics[width=\columnwidth, clip, trim= 0.0mm 0.0mm 0.0mm 0.0mm]{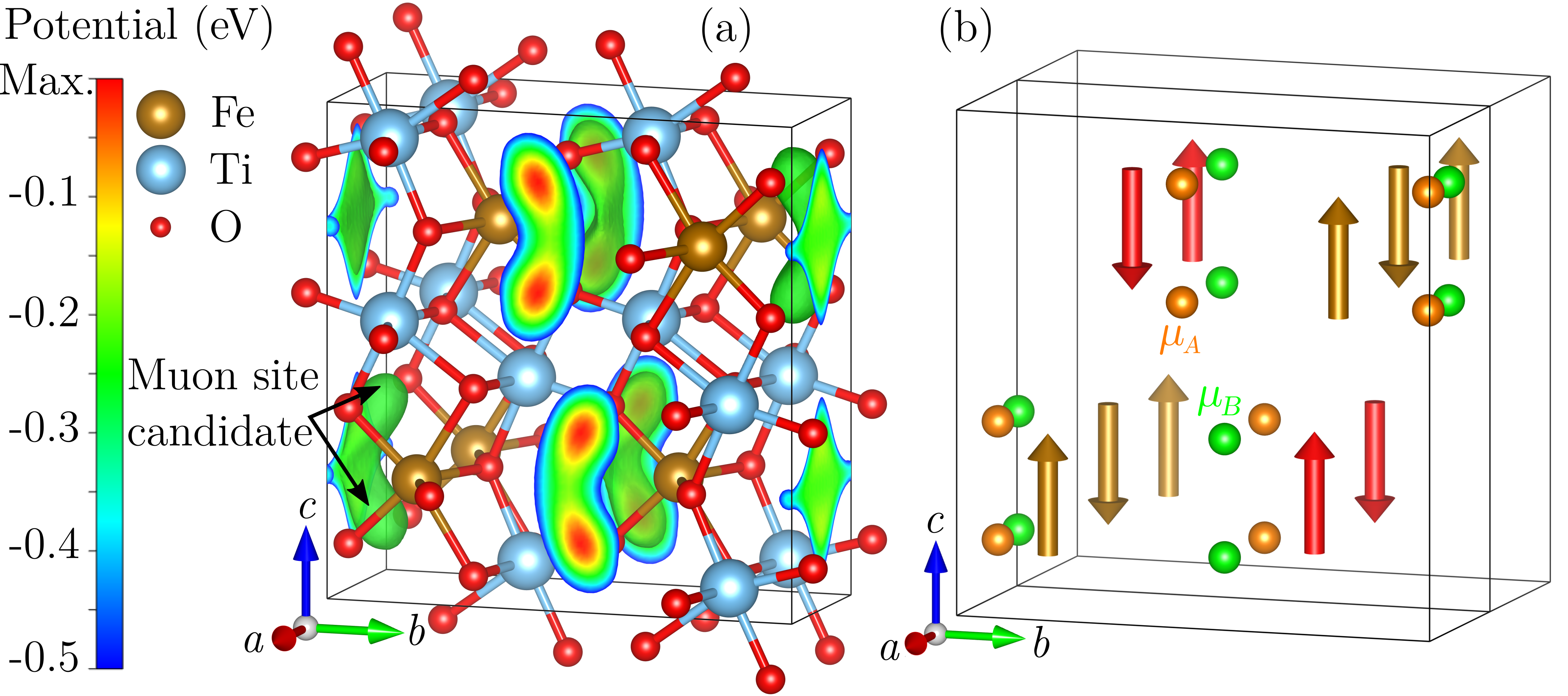}
\caption{ \label{plot:coulomb}
(a) Coulomb potential of \fto calculated via DFT. The green isosurface is plotted at $0.25$~eV below the maximum potential. (b) Muon stopping sites obtained through relaxation calculations, shown together with one of the magnetic domains found in \cto~\cite{Kirschner2019}. The 16 symmetry equivalent muon sites in the unit cell of the parent structure are shown, which map onto two groups $\mu_A$ and $\mu_B$ in the magnetic unit cell that are related by the mirror $\{m_x|0,0,0\}$. The Fe moments are shown for two unit cells of the parent structure to illustrate the AFM chains along $a$ depicted in brown and red. The visualisation software used is {\sc Vesta}~\cite{Momma2008}.
}
\end{figure}

We employed two complementary approaches to determine the potential muon stopping sites. First, we compute the electrostatic Coulomb potential from the converged electron density and plot it throughout the unit cell, as shown in Figure~\ref{plot:coulomb}(a). The maxima of the potential correspond to the lowest energies required to add a positive charge, and have been reliable indicators of possible muon sites~\cite{Moller2013_DFT}. From the electrostatic potential we can identify one muon site candidate at about \mbox{[0.45, 0.05, 0.15]}, which is only approximate as it does not account for local distortions caused by the implanted muon. In a second approach, we insert a single muon into the unit cell at various different positions and allow all ions to move until the structure finds its lowest energy state. These relaxation calculations include muon-induced perturbations and yield the energetically most favourable muon location to be at \mbox{[0.331, 0.045, 0.133]}. This site represents a simple distortion from the location of the maximum potential towards the nearest oxygen, with which the muon forms a 1.0~$\si{\angstrom}$ O--H like bond. Such bonds appear to be typical for muon sites in oxygen containing compounds~\cite{Moller2013_DFT,Foronda2015,Kirschner2018,Princep2019}. This fully relaxed site is also completely consistent with that found in \cto, although the local distortions it causes are subtly different. While the presence of the muon seems to perturb the O and Ti in much the same way in the two compounds, we find that our DFT+$\mu$ calculations predict a significantly reduced effect on the Fe ions. Whereas the Co ions are on average displaced by $\sim$0.12~$\si{\angstrom}$, the Fe ions only move by about $\sim$0.05~$\si{\angstrom}$. One possible explanation is that because the Fe ions have one fewer electron in the outer shell they are less strongly repulsed by the positive muon, although other factors beyond such a simple Coulomb-like picture may also play a role.

\section{Dipole field calculations}\label{sec:dipolefields}
The muon site calculated via DFT has 16 symmetry-equivalent position in the unit cell of the $Cmcm$ parent structure. However, we only observe two muon signals experimentally, which requires that the 16 crystallographically equivalent muon sites must map into two inequivalent groups of sites within the magnetically ordered state of \fto. Furthermore, the experimental signals have an amplitude ratio of about 1:1, which tells us that each group should contain 8 of the 16 symmetry-equivalent muon sites. This restricts the possible magnetic structure of \fto greatly. One possible candidate magnetic structure is that found in \cto~\cite{Kirschner2019}, in which the mirror $\{m_x|0,0,0\}$, which is broken at the transition observed in \cto, would be a natural operation for mapping the 16 muon sites onto two groups $\mu_A$ and $\mu_B$, as illustrated in Figure~\ref{plot:coulomb}(b). The magnetic fields at $\mu_A$ and $\mu_B$ would be swapped upon application of the $m_x$ operator. Assuming the propagation vector $\boldsymbol{k}$=(1/2,1/2,0) of one of the domains found in \cto, but allowing for a different Fe moment size and direction, we can compute the expected dipolar fields at the muon stopping sites. By comparing the number, relative sizes and multiplicities of the computed fields, presented in detail in the Appendix, we can constrain the possible moment directions and sizes to those summarised in Table~\ref{TAB:msize}. These give values for the relative error between observed and computed fields of 6~\% to 10~\%, showing a good agreement with our experimental data, particularly considering that we have neglected hyperfine fields. If one assumes an octahedral environment of the Fe$^{2+}$ and Co$^{2+}$ ions and high spin states we would expect the ratio of moment sizes to be $m_{\rm Fe}/m_{\rm Co}$=$\sqrt{6/(15/4)}$$\approx$1.265. This suggests that the Fe$^{2+}$ moment is most likely to lie along $c$ as this yields a moment size approximately 1.265 times the Co$^{2+}$ moment size reported for \cto.\cite{Kirschner2019} However, we cannot unequivocally assign this direction as the crystal environment is neither octahedral, nor should the $g$-factors of the Fe and Co ions necessarily be the same, nor is the site mixing the same in the two studied compounds. 

\begin{table}
\caption{\label{TAB:msize}Fe moment sizes that minimise disagreement between experimental and calculated fields strengths for different moment directions. The figure of merit is taken as $\chi$=$\sqrt{((B_1-B_{1}^{\rm exp})/B_{1}^{\rm exp})^2+((B_2-B_{2}^{\rm exp})/B_2^{\rm exp})^2}$ and is a measure of the relative error between the measured and calculated fields.}
\begin{ruledtabular}
\begin{tabular}{ c c c c c}
$\hat{\boldsymbol{m}}_{\rm Fe}$ & $|\boldsymbol{m}_{\rm Fe}|$ ($\mu_{\rm B})$ & $B_1$ (T) & $B_2$ (T)& $\chi$ (\%)\\
\hline
$c$ & 3.44 & 0.382 & 0.443 & 6.3 \\
$a$ & 2.54 & 0.382 & 0.447 & 5.9 \\
$b$ & 2.86 & 0.392 & 0.431 & 10.1 \\
$\frac{a+b}{2}$ & 2.57 & 0.386 & 0.439 & 7.8 \\
\multicolumn{2}{l}{experimental} & \phantom{(4)}0.367(4) & \phantom{(5)}0.466(5) &
\end{tabular}
\end{ruledtabular}
\end{table}

\section{Conclusions\label{sec:conclusion}}
We have observed long-range magnetic order in \fto, which is consistent with the antiferromagnetic state previously reported in \cto, but with a substantially higher transition temperature of 41.8(5)~K. We presented detailed calculations of the muon site candidates and dipolar fields at these sites for different moment directions and sizes. While a small number of moment directions was found to give fields consistent with the experimentally observed ones, we argued that the most likely direction is along $c$ with a moment size of $m_{\rm Fe}$=3.44~$\mu_{\rm B}$. Furthermore, our x-ray diffraction data show \fto to have the same crystal structure as \cto. Given that these two compounds are therefore isostructural crystallographically and magnetically this means that \fto is most likely also a spin Jahn-Teller driven antiferromagnet like \cto. This leads to the interesting question whether similar behaviour may be common to related orthorhombic systems. In addition, we find there is significant missing entropy associated with the magnetic heat capacity, which indicates strong correlations above the transition. One possible explanation is the formation of antiferromagnetic correlations along the $a$ direction (the most strongly coupled direction because of Fe--O--Fe superexchange) persisting above $T_{\rm N}$ (but below $T$$\sim$$J/k_{\rm B}$). The amount of site mixing is another important aspect to consider. A larger site mixing would effectively reduce the magnetic contribution to the free energy, and thereby might reduce the transition temperature or suppress the transition altogether. A future study in which the site mixing is increased purposefully could study the effect of this on the magnetic ordering. Furthermore, it would be useful to study the effect of increasing the cation size even further, especially since the orthorhombic crystal structure may become unstable for larger cations, similar to the Goldschmidt tolerance factor in perovskites~\cite{Goldschmidt1926,Li2016}.

\begin{acknowledgments}
This work is supported by EPSRC (UK) grant No EP/N024028/1. R.D.J. acknowledges support from a Royal Society University Research Fellowship. Part of this work was performed at the Swiss Muon Source, PSI Switzerland. The authors made use of the University of Oxford Advanced Research Computing (ARC) facility~\cite{ARC}.
\end{acknowledgments}

\begin{appendices}
\section{$m_{\rm Fe}$ direction and size determination}
Denoting the Fe moments by $\boldsymbol{m}_i$ at positions $\boldsymbol{r}_{i}$ we can calculate the dipolar fields at the muon site via
\begin{equation}
\boldsymbol{B}_\mathrm{dip}(\boldsymbol{r}_\mu)=\frac{\mu_0}{4\pi}\sum_i \frac{3(\boldsymbol{m}_i\cdot\hat{\boldsymbol{r}}_{i\mu})\hat{\boldsymbol{r}}_{i\mu}-\boldsymbol{m}_i}{|\boldsymbol{r}_{i\mu}|^3}
,\end{equation}
with $\boldsymbol{r}_{i\mu}$=$\boldsymbol{r}_i$$-$$\boldsymbol{r}_\mu$ and $\boldsymbol{r}_\mu$ denoting the position of the muon. The possible muon positions are taken to be the 16 symmetry-equivalent sites per crystallographic unit cell that correspond to the most likely muon site candiate calculated from DFT. The positions of the Fe ions are taken to be those determined by our x-ray diffraction experiments, and we refer to the four symmetry-equivalent crystallographic sites with respect to the $Cmcm$ unit cell as Fe1: [0,$y$,$\frac{1}{4}$], Fe2: [$\frac{1}{2}$,$\frac{1}{2}$$-$$y$,$\frac{3}{4}$], Fe3: [$\frac{1}{2}$,$\frac{1}{2}$$+$$y$,$\frac{1}{4}$], Fe4: [0,1$-$$y$,$\frac{3}{4}$], where $y$=0.19322(6). Because all the Fe ions are structurally equivalent and hence share the same chemical environment, we furthermore constrain the magnetic moment sizes of these ions to be equal. The propagation vector is taken to be $\boldsymbol{k}$=(1/2,1/2,0), corresponding to one of the domains found in \cto, as it is the most likely candidate given the similarity between the crystal structures and the observed muon spectra of \fto and \cto. Within these assumptions we performed a symmetry analysis using the {\sc isotropy} Suite~\cite{Campbell2006, Stokes2007} taking the $Cmcm$ crystal structure as the parent. Through systematic tests it was found that only two scenarios exist that lead to magnetic structure models with a single irreducible representation. These two scenarios area: \RomanNumeralCaps{1}) the moments on sites Fe1 and Fe4 are parallel, and moments on the Fe2 and Fe3 sites are antiparallel. If the moment is along $c$ the relevant irrep is mS$^-_2$ (space group $P_a2_1/m$, this is the structure reported for \cto)\cite{Kirschner2019}, and if the moment is perpendicular to $c$ the irrep is mS$^-_1$ (space group $P_c2_1/c$). \RomanNumeralCaps{2}) the moments on sites Fe1 and Fe4 are antiparallel, and moments on the Fe2 and Fe3 sites are parallel. If the moment is along $c$ the relevant irrep is mS$^+_2$ (space group $P_a2_1/m$), and if the moment is perpendicular to $c$ the irrep is mS$^-_1$ (space group $P_c2_1/c$). In both of these scenarios two domains exist, with the moments on sites Fe1 and Fe2 either aligned parallel or antiparallel to each other. Both domains yield the same dipolar fields at the muon sites and hence cannot be distinguished by our measurements. 

\begin{figure}[htb] 
\center
\includegraphics[width=\columnwidth, clip, trim= 0.0mm 0.0mm 0.0mm 0.0mm]{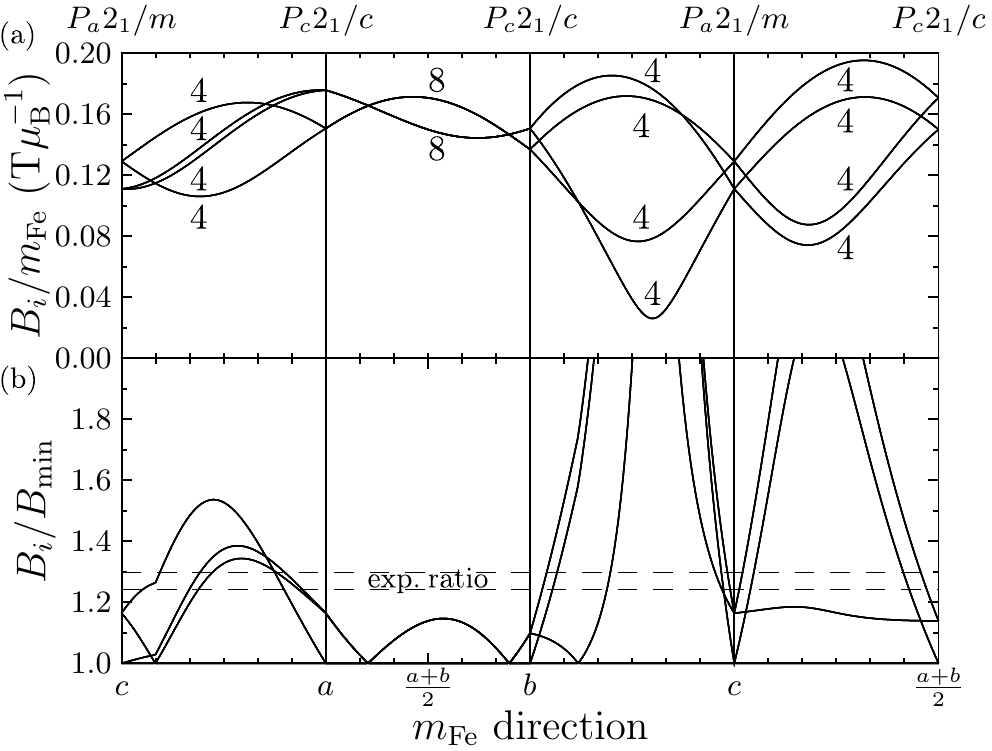}
\vspace{-20pt}
\caption{ \label{plot:mdirection}
Dipolar fields $B_i$ at DFT calculated muon sites as a function of Fe moment direction. The propagation vector was taken to be $\boldsymbol{k}$=(1/2,1/2,0) consistent with that of \cto, with all the $m_{\rm Fe}$ the same size, an example of which is shown in Figure~\ref{plot:coulomb}(b), and the moments on the Fe1 and Fe4 sites were taken to be parallel (see the description of scenario \RomanNumeralCaps{1})  above). The magnetic space groups for different directions are given at the top. (a) Calculated magnetic fields normalised to the Fe moment size. The numbers next to the curves represent the multiplicity of the curves, i.e. how many of the 16 symmetry-equivalent muons in the crystallographic unit cell would experience this field. (b) Calculated magnetic fields $B_i$ divided by the smallest computed field $B_{\rm min}$=min($B_i$), which corresponds to the curve of lowest field in panel (a) and is guaranteed to be present in our data. The two horizontal dashed lines represent the upper and lower bounds of the ratio of the two experimentally observed fields.
}
\end{figure}

\begin{figure}[htb] 
\center
\includegraphics[width=\columnwidth, clip, trim= 0.0mm 0.0mm 0.0mm 0.0mm]{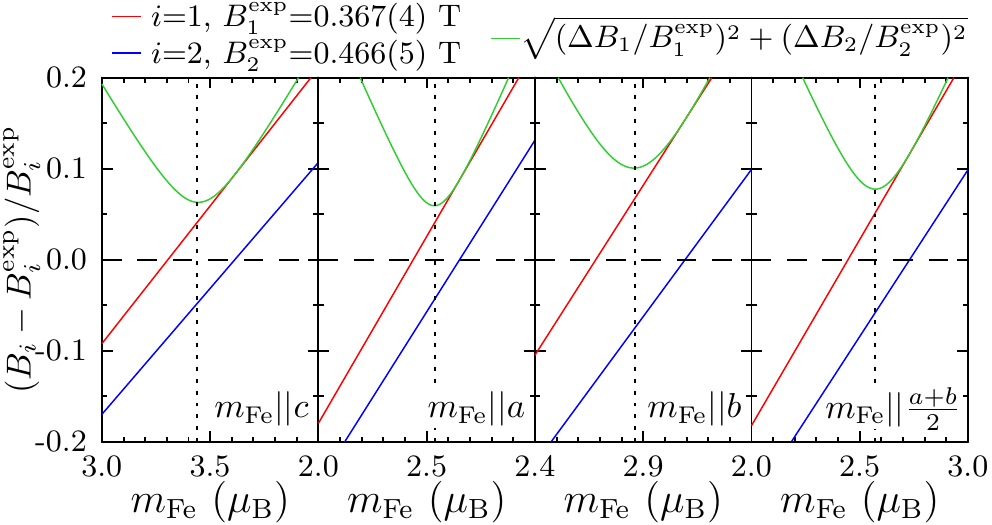}
\vspace{-20pt}
\caption{ \label{plot:msize}
Dipolar fields at DFT calculated muon sites as a function of Fe moment size for a selected number of Fe moment directions for scenario \RomanNumeralCaps{1}) as described in the main body. The vertical dotted lines indicates the values of $m_{\rm Fe}$ that minimise $\sqrt{(\Delta B_1/B_{1}^{\rm exp})^2+(\Delta B_2/B_{2}^{\rm exp})^2}$, where $\Delta B_i=B_i-B_i^{\rm exp}$, and the $B_i^{\rm exp}$ are the low temperature fields observed in our \musr experiments.
}
\end{figure}

In Figure~\ref{plot:mdirection} we plot the dipolar fields obtained for a range of possible Fe moment directions for scenario \RomanNumeralCaps{1}) as described above. We see that for the vast majority of tested directions we obtain fields that are inconsistent with the experimentally observed ones. (We can probably accept a value of $B_i$/$B_{\rm min}$ which is a little outside the experimentally determined ratio of 1.27(3) shown as dashed lines in Figure~\ref{plot:mdirection} because of the uncertainty due to the unknown value of the hperfine fields). This is because either these moment directions yield more than two fields, or two fields with multiplicity ratios that are far from the experimental ratio of about 1:1, or because the ratio of the two field strengths is vastly different from that of the experimental one. This allows us to restrict the possible moment directions to be along, or very close to, one of the orthorhombic axes or along $(a+b)$/2. For these directions we obtain two local fields with ratios for the multiplicity and strength consistent with our data. We can scale these fields with the moment size $m_{\rm Fe}$, as shown in Figure~\ref{plot:msize}, and minimise the value of $\chi$=$\sqrt{((B_1-B_{1}^{\rm exp})/B_{1}^{\rm exp})^2+((B_2-B_{2}^{\rm exp})/B_2^{\rm exp})^2}$ to optimise the agreement with our experimental observations. This leads to potential magnetic structures with moment sizes and directions summarised in Table~\ref{TAB:msize}. 

\begin{figure}[htb] 
\center
\includegraphics[width=\columnwidth, clip, trim= 0.0mm 0.0mm 0.0mm 0.0mm]{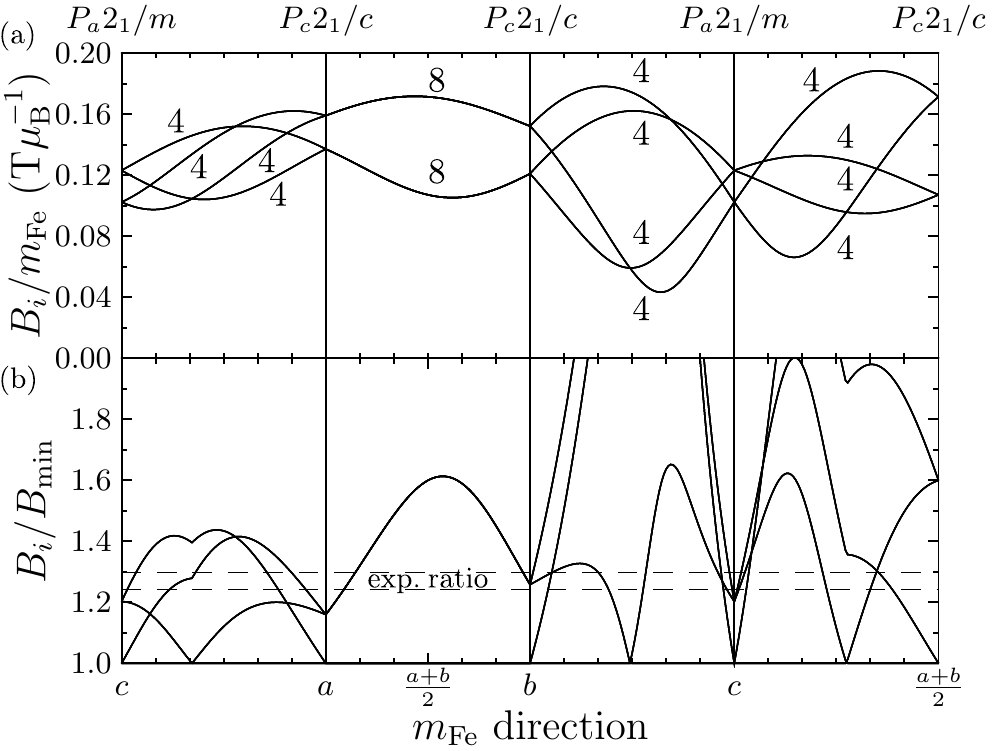}
\vspace{-20pt}
\caption{ \label{plot:mdirection2}
As Figure~\ref{plot:mdirection} but with moments on the Fe1 and Fe4 sites aligned antiparallel in line with scenario \RomanNumeralCaps{2}) above.
}
\end{figure}

\begin{figure}[htb] 
\center
\includegraphics[width=\columnwidth, clip, trim= 0.0mm 0.0mm 0.0mm 0.0mm]{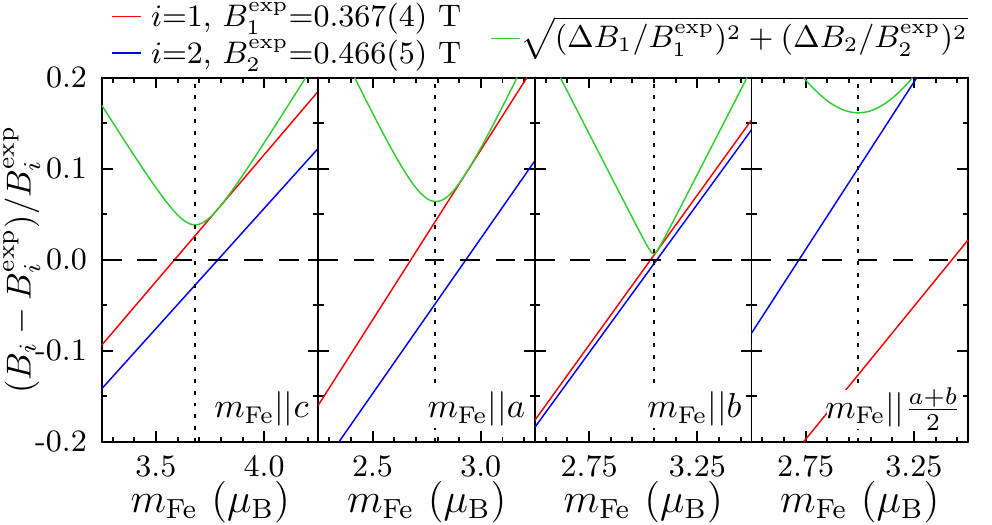}
\vspace{-20pt}
\caption{ \label{plot:msize2}
As Figure~\ref{plot:msize} but with moments on the Fe1 and Fe4 sites aligned antiparallel in line with scenario \RomanNumeralCaps{2}) above.
}
\end{figure}

For completeness we have repeated these calculations for the case of the moments on the Fe1 and Fe4 sites antiparallel, and the moments on the Fe2 and Fe3 sites parallel, which is referred to as scenario \RomanNumeralCaps{2}) above. The results are given in Figures~\ref{plot:mdirection2} and~\ref{plot:msize}, and Table~\ref{TAB:msize2}. They show that similar to scenario \RomanNumeralCaps{1}) only a few moment directions would be consistent with our experimental data (two fields with the correct ratios of field strengths and multiplicities). While some of these directions can lead to good agreement with the experimentally observed fields, the moment size predicted within these magnetic structures differ significantly from those one expects for the Fe$^{2+}$ ions: the Co moment in \cto was measured to be 2.72~$\mu_{\rm B}$ and a simple calculation based on octahedral environments and high spin states of the ions yields $m_{\rm Fe}$=$\sqrt{6/(15/4)}$$m_{\rm Co}$$\approx$3.44~$\mu_{\rm B}$. Therefore, we believe the most likely magnetic structure of \fto that is consistent with our data to be isostructural to that reported for \cto:\cite{Kirschner2019} Fe moments of equal magnitude aligned along $c$ with their relative alignment as described in scenario \RomanNumeralCaps{1}) above (magnetic space group $P_a2_1/m$). However, given the inherent uncertainties of our calculations (we have neglected hyperfine field contributions, the environments of the ions are not octahedral, the effect of site mixing has not been quantified) we cannot conclusively rule out that \fto might adopt some of the other tested magnetic structures that we have tested. We note that the case of \RomanNumeralCaps{2}) with $m_{\rm Fe}$$\parallel$$c$ (magnetic space group $P_a2_1/m$) would also lead to a spin Jahn-Teller driven transition due to the complete frustration of the exchange pathways between the ions. 

\begin{table}
\caption{\label{TAB:msize2}As Table~\ref{TAB:msize} but with moments on the Fe1 and Fe4 sites aligned antiparallel in line with scenario \RomanNumeralCaps{2}) above.}
\begin{ruledtabular}
\begin{tabular}{ c c c c c}
$\hat{\boldsymbol{m}}_{\rm Fe}$ & $|\boldsymbol{m}_{\rm Fe}|$ ($\mu_{\rm B})$ & $B_1$ (T) & $B_2$ (T)& $\chi$ (\%)\\
\hline
$c$ & 3.68 & 0.377 & 0.453 & 3.8 \\
$a$ & 2.79 & 0.383 & 0.443 & 6.4 \\
$b$ & 3.05 & 0.369 & 0.464 & 0.7 \\
$\frac{a+b}{2}$ & 2.99 & 0.320 & 0.512 & 16\\
$\cos(10\si{\degree})a$+ &\multirow{2}{*}{2.83} & \multirow{2}{*}{0.368} & \multirow{2}{*}{0.465} & \multirow{2}{*}{0.3}\\
$\sin(10\si{\degree})b$\phantom{+} &\\
\multicolumn{2}{l}{experimental} & \phantom{(4)}0.367(4) & \phantom{(5)}0.466(5) &
\end{tabular}
\end{ruledtabular}
\end{table}

\end{appendices}
\FloatBarrier
\bibliographystyle{apsrev4-1}
\bibliography{FeTi2O5}
\end{document}